\documentclass[twocolumn,color,superscriptaddress,psfig,showpacs,amsmath,amssymb,floatfix,fleqn]{revtex4-1}
\usepackage{graphicx}
\usepackage{subfigure}
\usepackage{leftidx}
\usepackage{amsmath}
\usepackage{float}
\usepackage{color}
\usepackage[misc]{ifsym}
\usepackage{hyperref}
\hypersetup{
            linkcolor=blue,
            anchorcolor=blue,
            citecolor=blue,
            urlcolor=blue,
}

\usepackage{bookmark}
\usepackage{silence}
\DeclareMathOperator{\sech}{sech}

\usepackage{silence}
\begin{document}


\title{Nonlinear dynamics in the flexible shaft rotating-lifting system of silicon crystal puller using Czochralski method}

\author{Hai-Peng Ren}
 \email{renhaipeng@xaut.edu.cn}
\affiliation{
Shaanxi Key Laboratory of Complex System Control and Intelligent Information Processing, Xi'an University of Technology, Xi'an, Shaanxi 710048, People’s Republic of China
}%
\author{Zi-Xuan Zhou}%
\affiliation{
Shaanxi Key Laboratory of Complex System Control and Intelligent Information Processing, Xi'an University of Technology, Xi'an, Shaanxi 710048, People’s Republic of China
}%
\affiliation{%
North Minzu University Library, Yinchuan, Ningxia 750021, People’s Republic of China
}%

\author{Celso Grebogi}
\affiliation{
Shaanxi Key Laboratory of Complex System Control and Intelligent Information Processing, Xi'an University of Technology, Xi'an, Shaanxi 710048, People’s Republic of China
}%
\affiliation{%
Institute for Complex System and Mathematical Biology, University of Aberdeen AB24 3UE, United Kingdom
}%




\date{\today}

\begin{abstract}
Silicon crystal puller (SCP) is a key equipment in silicon wafer manufacture, which is, in turn, the base material for the most currently used integrated circuit (IC) chips. With the development of the techniques, the demand for longer mono-silicon crystal rod with larger diameter is continuously increasing in order to reduce the manufacture time and the price of the wafer. This demand calls for larger SCP with increasing height, however, it causes serious swing phenomenon of the crystal seed. The strong swing of the seed causes difficulty in the solidification and increases the risk of mono-silicon growth failure.The main aim of this paper is to analyze the nonlinear dynamics in the FSRL system of the SCP. A mathematical model for the swing motion of the FSRL system is derived.  The influence of relevant parameters, such as system damping, excitation amplitude and rotation speed, on the stability and the responses of the system are analyzed. The stability of the equilibrium, bifurcation and chaotic motion are demonstrated, which are often observed in practical situations. Melnikov method is used to derive the possible parameter region that leads to chaotic motion. Three routes to chaos are identified in the FSRL system, including period doubling, symmetry-breaking bifurcation and interior crisis. The work in this paper explains the complex dynamics in the FSRL system of the SCP, which will be helpful for the SCP designers in order to avoid the swing phenomenon in the SCP.
\end{abstract}


\maketitle


\section{Introduction}

As the main material base for IC chip production, mono-silicon wafer production plays an important role in modern industrial field. The mono-silicon wafer is made from the mono-silicon rod produced by silicon crystal puller using the Czochralski (Cz) method [1]. In the Cz method, the polycrystalline silicon blocks are put into a crucible and melted by a heater surrounding the crucible at about $1420\,^{\circ}\mathrm{C}$. A mono-silicon seed  hanged at the end of the flexible shaft rotating-lifting system is dropped into the melting silicon, provided the proper conditions are obeyed. As the flexible shaft rotating counterclockwise and the crucible rotating clockwise, the mono-silicon seed is slowly lifted upward to allow the new crystal growth. By precisely controlling the temperature gradients and the rate of lifting, a mono-silicon crystal ingot is extracted from the melt. During the whole procedure of the mono-silicon rod production, the FSRL system rotates and lifts the crystal rod at a certain rate determined by the technique parameters. The rotation of the mono-silicon crystal seed mixes the silicon melt and makes the crystal/melt surface to have a radius uniformity, which is essential for the quality of the mono-silicon crystal [2-4]. There is an increasing demand for longer mono-silicon crystal rods with increasing diameter in order to reduce the manufacture time and to improve the utilization rate of the wafer. Larger SCPs with increasing height being put into usage lead to stronger swing phenomenon of the crystal seed. Specifically, in the seeding stage of crystal growth process, the swing phenomenon increases the risk of mono-silicon growth failure or causes defects in the growth of the silicon crystal. The engineering observation is that the swing amplitude and frequency suddenly become irregular under some circumstances. The SCP operator usually adjusts the rotation speed to avoid such unexpected irregular swing. But with the larger SCP size, this unexpected phenomenon becomes more frequent with even larger amplitude. How to characterize this phenomenon from a dynamical system viewpoint is of practical significance in the engineering field.

\begin{figure*}[!htb]
  \centering
  \subfigure[]{
    \includegraphics[width=2in]{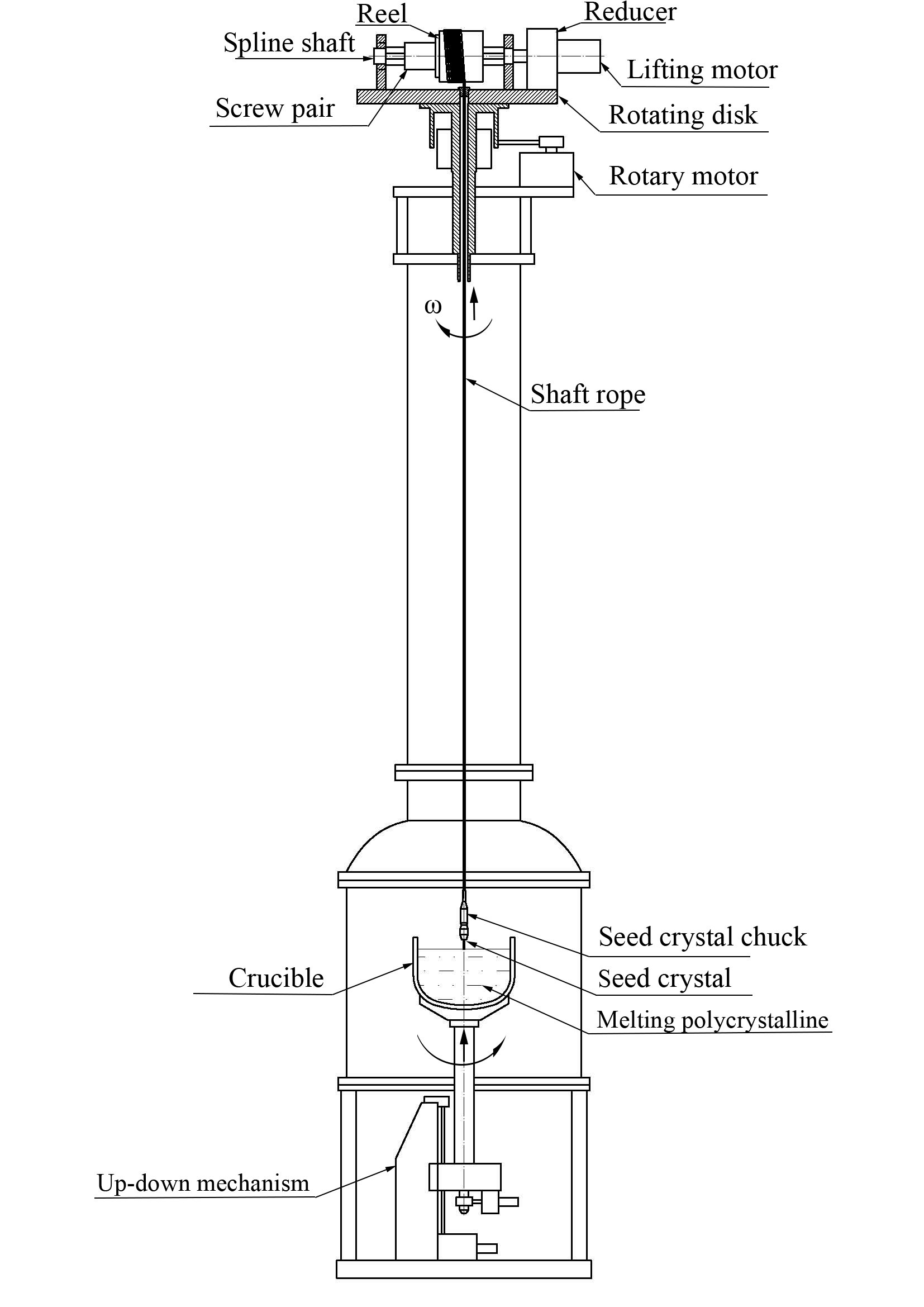}
  }
  \subfigure[]{
  \includegraphics[width=2in]{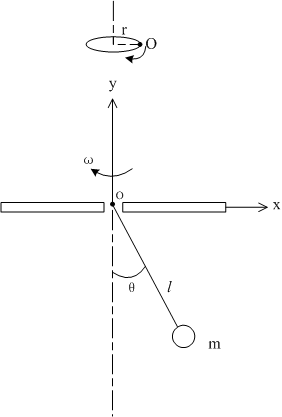}
  }
\caption{(a) Structure diagram of flexible shaft rotating-lifting system of the Czochralski silicon crystal puller, and (b) its simplified model.}
\label{fig1}

\end{figure*}

Up to now, few works have considered the dynamics of the swing phenomenon in the SCP. Yuan assumed the FSRL system to be like a double pendulum, and studied the relationship of the swing amplitude and the rotation speed (frequency) [5]. However, there are two weak points in that work: first, it is unreasonable to treat the FSRL system as a double pendulum, especially, at the initial stage of the mono-silicon rod growth from the melt; second, only simple oscillation is observed from the model without the systematical analysis of the whole dynamics. In a subsequent work [6], Yuan established a four-degree of freedom nonlinear dynamic equations by considering the in-plane and out-plane vibrations of the FSRL system. Then it deduced a linear approximation model of the system. Numerical simulations are given to show that the oscillation could be diminished by reducing the error of centration and by increasing the damping. However, it is also unreasonable to analyze the FSRL system by using linear models, especially to address the oscillation. Moreover, the damping between the solid mono-silicon crystal seed and the liquid polycrystal melt is very small, and it cannot be increased or decreased. The irregular swing phenomenon and the underlying dynamics are still unclear, which need to be further and deeper investigated.

This paper is organized as follows. Section 2 describes the structure and working principle of the system. Section 3 introduces the mathematical modeling of the system. Section 4 uses Melnikov method to obtain the parameter region where chaos might exist. Section 5 studies the stability and bifurcation of the unperturbed system. Section 5 presents the numerical simulation to show the dynamic response of the system with perturbation, bifurcation diagrams, the Lyapunov exponents, phase trajectories, Poincar\'{e} sections, and power spectrum. Three routes to chaos are analyzed in section 6. Finally, section 7 summarizes the main results and the contributions of this article.

\section{The system configuration and working principle}

The simplified structure diagram of SCP is given in Fig. 1a. From Fig. 1a, the puller consists of four parts, including the base pedestal usually placed underground to support the whole puller upside and the crucible up-down mechanism, the main body of the crucible and heater inside, the puller neck to hold the long crystal ingot rod, and the head with the rotating-lifting mechanism. The flexible shaft rope is curled around the reel mounted on the spline shaft driven by the lifting motor through the reducer. The lifting motor regulates the lifting rate of the crystal ingot rod. A screw pair on the spline shaft is used to make the rotation shaft rope to be located at the center. All the lifting elements are installed on the rotating disk, which is driven by the rotating motor through the reducer. The rotation of the rotating disk drives the flexible shaft rotation around the center.

From the above description, we learn that the FSRL system can be treated as a pendulum with moving pivot, as shown in Fig. 1b. Works on parametrically excited pendulum has been reported in [7-12]. The models in those papers are usually abstracted from various actual mechanical devices, such as the mechanical components [13, 14], rotary cranes [15, 16], and energy converters [17]. Researches have shown that, a mathematical model like a rotating pendulum exhibit chaotic phenomena [18, 19]. The pendulum model in this paper, however, is different from the traditional parametric pendulum. Due to the imperfection of the manufacture, the rotating disk might have eccentricity, which makes the suspension point periodically varying. The way the suspension point $O$ moves can be illustrated by the upper part in Fig. 1b. The period is decided by the rotating disk rotation frequency. In our model process, there is no linearization is considered, which reveal the nature of the nonlinear dynamics.

In order to explain the swing phenomenon and to understand the dynamics of the FSRL system, we establish the mathematical model of this system and analyze the dynamical characteristics of the FSRL system. The main purpose of this work is to demonstrate that the FSRL system can generate different kinds of motion, from periodic oscillations to chaos, when the rotational frequency of the crystal is close to the natural frequency of the flexible shaft. We show that period doubling bifurcation, symmetry-breaking bifurcation and interior crisis can be present in the FSRL system. A better understanding of the FSRL system dynamics will help engineers to control the swing in an effective and efficient way in order to ensure a proper stable crystal growth environment.

\section{Mathematical model of the FSRL system in SCP}
In this paper, we focus on the model of the FSRL system in the crystal seeding stage, in which the crystal seed (about 10 mm in diameter) can be treated as a mass point. Different from the previous double pendulum or 4 freedom oscillation equation, a pendulum with a moving suspension point is the feature of our model, where the moving of the suspension point is caused by the eccentricity of the rotating disk with respect to the center of the whole system.

The model is derived under the following three assumptions:

1. As the lifting speed is extremely slow with respect to the rotation, the length of the suspended flexible shaft can be treated as a constant.

2. The mass of the flexible shaft is neglected; the mass of the crystal chucks and (seed) crystal is assumed as a mass point.

3. Within the SCP, it is near vacuum state. The air damping of the system is too small to affect the system. The damping of the system is mainly caused by the inter action between the solid mono-silicon crystal rod(seed) and the polycrystalline silicon melt.

The simplified diagram of the FSRL system of SCP is shown in Fig. 1b. The rotation motors drives the rotating disk with angular velocity $\omega$. The flexible shaft length is \textit{l}, and the seed crystal together with ingot crystal has a mass \textit{m}.

The system is considered to be a rotating pendulum, and the general nonlinear differential equations can be derived by using the second kind Lagrange's equation.


Define the angle between the rotational axis and the flexible shaft as the generalized coordinate, $\theta$, as shown in Fig. 1b. The level of the rotating disk is assumed to be the zero potential energy surface. Then, the kinetic energy \emph{T} and the potential energy \emph{V} of the system are written as follows:
\begin{equation}
\nonumber
  T=\frac{1}{2}m(l^2\theta^2+l^2\omega^2\sin^2\theta),
\end{equation}
\begin{equation}
\nonumber
  V=-mgl\cos\theta.
\end{equation}
The Lagrangian of the system is, then,
\begin{equation}
\nonumber
  L=T-V=\frac{1}{2}m(l^2\theta^2+l^2\omega^2\sin^2\theta)+mgl\cos\theta.
\end{equation}
The periodic perturbed force caused by the eccentricity is given as:
\begin{equation}
\nonumber
  Q_F=mr\omega^2\cos(\omega t).
\end{equation}

In the practical system, \emph{r} is the eccentric distance. Using the Lagrange's equation, the dynamic equation of the rotating pendulum can be given as:
\begin{equation}%
  \ddot{\theta}=\frac{r}{l}\omega^2\cos(\omega t)+\omega^2\sin\theta\cos\theta-\frac{g}{l}\sin\theta-\frac{\xi}{m}\dot{\theta},
\end{equation}
where $\xi$ is the damping coefficient. Introducing dimensionless time $\tau=\omega_0t$, where $\omega_0=\sqrt{g/l}$ is the natural frequency of the pendulum, and then the dimensionless coordinates $\theta\equiv\theta$, we have the dimensionless form of the dynamics as follows:
\begin{equation}%
  \ddot{\theta}=A\Omega^2\cos(\Omega\tau)+\Omega^2\sin\theta\cos\theta-\sin\theta-c\dot{\theta},
\end{equation}
where $\Omega=\frac{\omega}{\omega_0}$, $A=\frac{r}{l}$, and $c=\frac{\xi}{m\omega_0}$.

Equation. (2) can be rewritten as state space equations: 
  \begin{align}
  \dot{x_1}&=x_2 \nonumber \\
  \dot{x_2}&=A\Omega^2\cos(\Omega\tau)+\Omega^2\sin{x_1}\cos{x_1}-\sin{x_1}-cx_2,
  \end{align}
where $x_1=\theta$ and $x_2=\dot{\theta}$. The dynamics of the flexible shaft rotating-lifting system is a two-dimensional non-autonomous system.

The phenomenon obtained by our model method is more reasonable to explain the practical observation, the on-going practice to control the swing also testifies the effectiveness of the model.

\section{Analysis of the unperturbed system}
The system without damping and perturbation is given by
  \begin{align}
  \dot{x_1}&=x_2 \nonumber \\
  \dot{x_2}&=\Omega^2\sin{x_1}\cos{x_1}-\sin{x_1}.
  \end{align}
System (4) is a Hamiltonian system and the Hamiltonian function is given by:
\begin{align}
  H(x_1,x_2)=\frac{1}{2}x_2^2+\frac{1}{4}\Omega^2\cos2x_1-\cos{x_1}.
\end{align}

\begin{figure}[!ht]
\centering
  \includegraphics[width=3.25in]{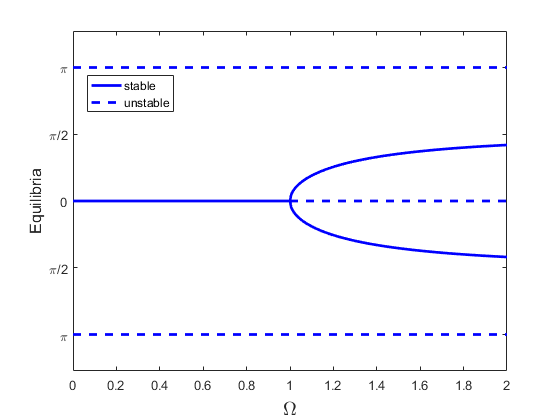}
\caption{Equilibria bifurcation diagram of system (4), where a pitchfork bifurcation is found, with stable equilibria given by \emph{solid line} and unstable equilibria given by \emph{dotted line}.}
\end{figure}

By analyzing the fixed points of system (4) and their stabilities, we obtain the following results:

(i) For $\Omega<1$, there is only one equilibrium O(0,0), being the center.

(ii) For $\Omega>1$, there are three equilibria including O(0,0), being the saddle, $C_1(x_0,0)$ and $C_2(-x_0,0)$ being the centers, where $x_0$ is the positive root of $x_1$ satisfying $\Omega^2\sin{x_1}\cos{x_1}-\sin{x_1}=0$.

System (4) undergoes pitchfork bifurcation at $\Omega=1$. The equilibria bifurcation diagram is given in Fig. 2. In addtion, with the help of the Hamilton function (5), the trajectories can be classified by the different values of the Hamiltonian $H(x_1,x_2)=E$, which are marked in the corresponding phase portraits shown in Fig. 3. It can be seen that the phase structure of system (4) will change according to parameter $\Omega$. In the case of $\Omega<1$, the orbits for $E<E_0$ $(E_0=2+\dfrac{1}{4}\Omega^2\cos2x_1-\cos{x_1})$ are represented by a family of ellipses, it means the system moves periodically around the minimum of potential energy, as shown in Fig. 3a. When $\Omega>1$, the phase portraits suddenly change into another structure, a pair of homoclinic orbits $q_+^0(t)$ and $q_-^0(t)$ connecting the origin to itself appear, plotted using red and blue solid lines, as shown in Fig. 3b, in the interior region of $q_+^0(t)$ and $q_-^0(t)$, there exists a family of periodic orbits.

\begin{figure*}
\subfigure{
  \centering
  \includegraphics[width=3.25in]{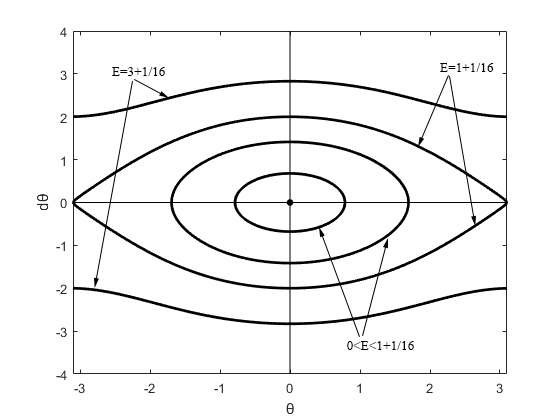}
}
\subfigure{
  \centering
  \includegraphics[width=3.25in]{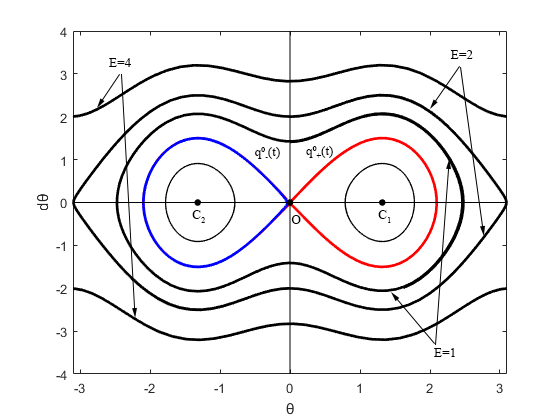}
}
\caption{(a) The phase portraits of system (2) for $\Omega=0.5$, (b) the phase portraits of system (2) for $\Omega=2$.}
\end{figure*}

Analytical expressions for the unperturbed homoclinic orbits can be derived by using Hamilton function (5). Notice that the solution of homoclinic orbits should satisfy the initial condition $(x_1(0),x_2(0)=(0,0))$, and then $\left .H(x_1,x_2)\right |_{(0,0)}=\dfrac{1}{4}\Omega^2-1$, we obtain:

\begin{equation}
  x_2^2=\dfrac{1}{2}\Omega^2-2-\dfrac{1}{2}\Omega^2\cos{2x_1}+2\cos{x_1},
\end{equation}

Equation (6) can be rewritten as follows:
\begin{align}
\nonumber
  \dfrac{dx_1}{dt}=\sqrt{\Omega^2-2-\Omega^2\cos^2x_1+2\cos{x_1}},
\end{align}
letting $\alpha^2=\Omega^2-1$ , it is rewritten as:
\begin{equation}
\nonumber
  dt=\dfrac{dx_1}{\sqrt{\alpha^2-1-\alpha^2\cos^2x_1-\cos^2x_1+2\cos{x_1}}}.
\end{equation}

Integrating both side of the above equation, we have:
\begin{equation}
\nonumber
  t=\pm\dfrac{1}{\alpha}\cosh^{-1}(\alpha\cot\dfrac{x_1}{2}).
\end{equation}

The above function can be transformed into:
\begin{equation}
\nonumber
  x_1(t)=\pm2\cot^{-1}(\dfrac{1}{\alpha})\cosh\alpha t.
\end{equation}

From $x_2(t)=\dfrac{dx_1(t)}{dt}$, we obtain the $x_2(t)$ in the follow form:
\begin{equation}
\nonumber
  x_2(t)=\mp\dfrac{2\alpha^2\sinh\alpha t}{\alpha^2+\cosh^2\alpha t}.
\end{equation}

We obtain the two homoclinic orbits:
\begin{equation}
  q^0_+(t)=(2\cot^{-1}(\dfrac{1}{\alpha})\cosh\alpha t,-\dfrac{2\alpha^2\sinh\alpha t}{\alpha^2+\cosh^2\alpha t}),
\end{equation}
and
\begin{equation}
  q^0_-(t)=(-2\cot^{-1}(\dfrac{1}{\alpha})\cosh\alpha t,\dfrac{2\alpha^2\sinh\alpha t}{\alpha^2+\cosh^2\alpha t}).
\end{equation}

The analytical expression of the homoclinic orbits of the unperturbed system obtained above enables us to investigate theoretically the chaotic motion in the original system.

\section{Parameter region of chaos existence using Melnikov method}
In this section, we will investigated the necessary condition for existing the chaotic motion in system (3) by using the Melnikov method. The Melnikov method is an analytical method to detect possible chaotic motion in Hamiltonian system. For a two-dimentional Hamiltonian system with the homoclinic or heteroclinic orbits, considering the perturbation of the system damping and periodic excitation, the distance between the stable and unstable manifolds of the system fixed point can be calculated by Melnikov’s integration. If the distance is equal to zero, the stable and unstable manifolds cross each other transversally, and from that crossing, the system will become chaotic [20].

We introduce the following notation for system (3):
\begin{equation}
  \boldsymbol{\dot{x}}=f(\boldsymbol{x})+g(\boldsymbol{x},t).
\end{equation}

Here $f(\boldsymbol{x})$ is the Hamiltonian system and $g(\boldsymbol{x},t)$ is the perturbation,
\begin{equation}
\nonumber
  f(\boldsymbol{x})=
  \begin{pmatrix}   
    x_2 \\  
    \sin x_1(\Omega^2\cos x_1-1)\\  
  \end{pmatrix}
  ,
\end{equation}
\begin{equation}
\nonumber
  g(\boldsymbol{x})=
  \begin{pmatrix}   
    0 \\  
    A\Omega^2\cos\Omega t-cx_2\\  
  \end{pmatrix}
  ,
    \boldsymbol{x}=
  \begin{pmatrix}   
    x_1 \\  
    x_2\\  
  \end{pmatrix}
  .
\end{equation}

Considering a Melnikov function defined as follows:
\begin{equation}%
  M(\tau)=\int_{-\infty}^{+\infty} \left[ f(q^0(t)) \wedge g(q^0(t),t+\tau) \right] \,dt
\end{equation}

where operation "$\wedge$" is defined as:
\begin{equation}%
\nonumber
  (a_1,a_2)^T \wedge (b_1,b_2)^T=a_1b_2-a_2b_1,
\end{equation}

Then, the Melnikov function $M(\tau)$ for the homoclinic orbits $q^0_+(t)$ of system (3) is given by:
\begin{align}%
\nonumber
  M(\tau) & =\int_{-\infty}^{+\infty} x_2(t)\left[ A\Omega^2\cos\Omega(t+\tau)-cx_2(t) \right] \,dt   \\
\nonumber
  &  =\int_{-\infty}^{+\infty} -\dfrac{2\alpha^2\sinh\alpha t}{\alpha^2+\cosh^2\alpha t}[A\Omega^2\cos\Omega(t+\tau)  \\
  &  -c\dfrac{-2\alpha^2\sinh\alpha t}{\alpha^2+\cosh^2\alpha t}] \,dt.
\end{align}

The computation for $q^0_-(t)$ can be conducted similarly. Since $x_2(t)$ is an odd function, equation (11) can be rewritten as:
\begin{align}%
\nonumber
  M(\tau) & =A\Omega^2\int_{-\infty}^{+\infty} \dfrac{2\alpha^2\sinh\alpha t}{\alpha^2+\cosh^2\alpha t}\sin\Omega t \,dt\sin\Omega\tau   \\
  &  -c\int_{-\infty}^{+\infty} (\dfrac{-2\alpha^2\sinh\alpha t}{\alpha^2+\cosh^2\alpha t})^2 \,dt.
\end{align}

The integrals in equation (12) can be calculated by:
  \begin{align}
  \nonumber
  I_1 & =\int_{-\infty}^{+\infty} \dfrac{2\alpha^2\sinh\alpha t}{\alpha^2+\cosh^2\alpha t}\sin\Omega t \,dt  \\
  \nonumber
  & =2\pi\sin [\dfrac{\Omega}{\alpha}\sinh^{-1}(\alpha)]\times\sech (\dfrac{\Omega\pi}{2\alpha}),
  \end{align}
  \begin{align}
  \nonumber
  & I_2 =\int_{-\infty}^{+\infty} (\dfrac{-2\alpha^2\sinh\alpha t}{\alpha^2+\cosh^2\alpha t})^2 \,dt  \\
  \nonumber
  & =4[\dfrac{\ln(\sqrt{\alpha^2+1}-\alpha)}{\sqrt{\alpha^2+1}}+\alpha],
  \end{align}

By calculating the above integrals, Melnikov function is given by:
\begin{align}
\nonumber
M(\tau) & =A\Omega^2\times 2\pi\sin [\dfrac{\Omega}{\alpha}\sinh^{-1}(\alpha)]\times\sech (\dfrac{\Omega\pi}{2\alpha})\times \sin{\Omega\tau}  \\
\nonumber
& -4c[\dfrac{\ln(\sqrt{\alpha^2+1}-\alpha)}{\sqrt{\alpha^2+1}}+\alpha].  \\
\end{align}

Melnikov's function (13) measures the distance between the stable and unstable manifolds in the Poincare section. If for all $\tau$ the following inequality (14) holds, the system might demonstrate chaotic behavior in the sense of Smale horseshoes.
\begin{equation}%
  \dfrac{A}{c}\ge\left | \dfrac{2[\dfrac{\ln(\sqrt{\alpha^2+1}-\alpha)}{\sqrt{\alpha^2+1}+\alpha}]}{\pi\Omega^2\sin [\dfrac{\Omega}{\alpha}\sinh^{-1}(\alpha)]\times\sech (\dfrac{\Omega\pi}{2\alpha})}\right |
\end{equation}
The condition given in (14) is consistent with the following result shown in Fig. 5.
%
%
%

\begin{figure*}
\subfigure{
  \centering
  \includegraphics[width=3.25in]{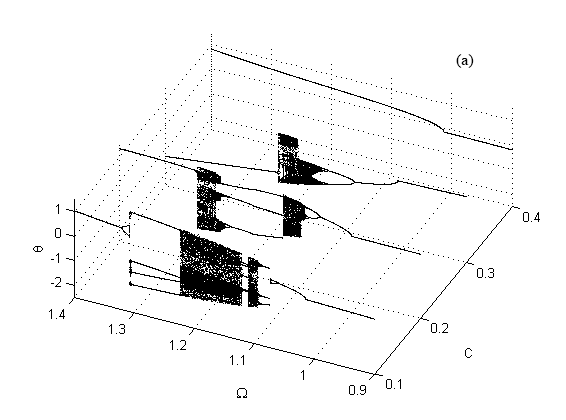}
}
\subfigure{
  \centering
  \includegraphics[width=3.25in]{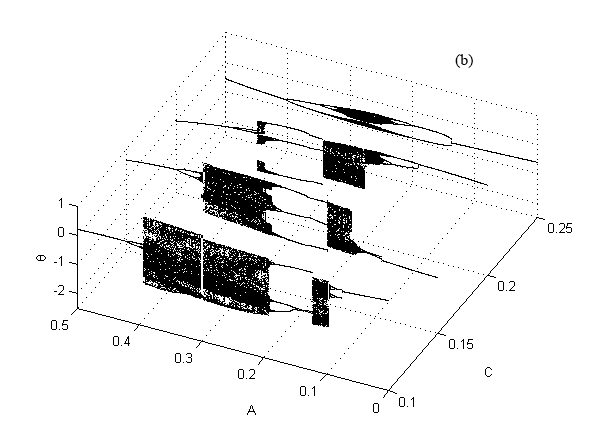}
}
\caption{The bifurcation diagrams of the system in three-dimensional space: (a) in $(c,\Omega,\theta)$ space for $A=0.2$, (b) in $(c,A,\theta)$ space for $\Omega=1.1$.}
\end{figure*}

\begin{figure}
\centering
  \includegraphics[width=3.5in]{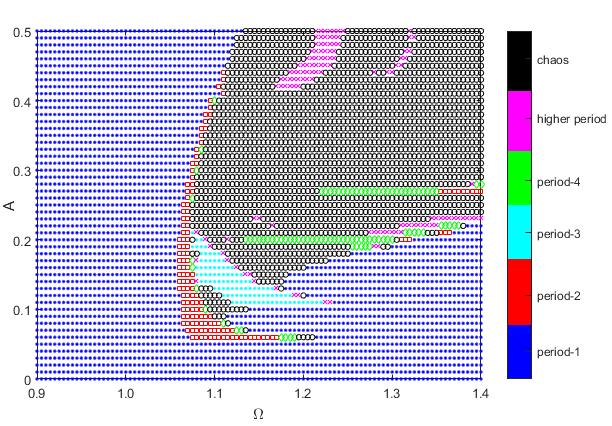}
\caption{Parameter space plot in the range of $\Omega\in(0.9,1.4)$ and $A\in(0,0.5)$, blue marks the period one motion, red marks the period two motion, green marks the period four motion, cyan marks the period three motion, magenta marks the higher period motion, and black marks the chaos.}
\end{figure}

\begin{figure*}

  \centering
  \includegraphics[width=3.25in]{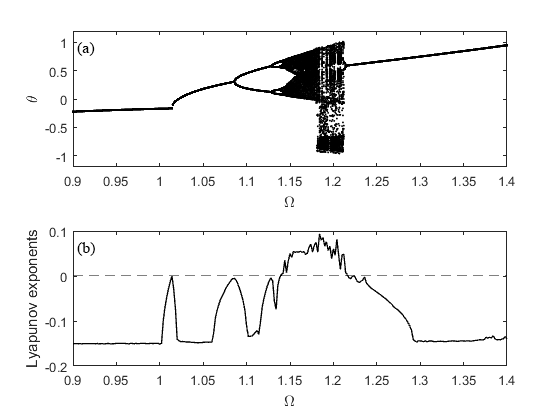}

\caption{(a) The bifurcation diagrams of the system for varying parameter $\Omega$, and (b) the LLE corresponding to the parameter range in(a).}
\end{figure*}
\begin{figure*}

  \centering
  \includegraphics[width=3.25in]{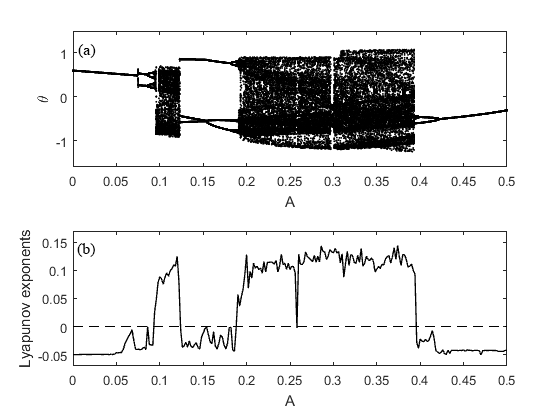}
\caption{(a) The bifurcation diagrams of the system for varying parameter $A$, and (b) the LLE corresponding to the parameter range in (a).}
\end{figure*}

\section{Dynamical behaviors of the FSRL system}

In order to investigate the dynamical behaviors of the full system (3), including the bifurcation diagrams, the Lyapunov exponents, and phase trajectories, Poincar\'{e} sections are used to show the complicated dynamics of system (3). Here, the fourth-order Runge-Kutta algorithm is used for the integration and the solution of the differential equations. 

\subsection{Bifurcation diagrams}
There are three parameters in system (3): the damping coefficient, the frequency and the amplitude of periodic excitation force caused by the eccentricity. The three-dimensional bifurcation diagrams of the system in space $(c,A,\theta)$ and $(c,\Omega,\theta)$ are given in Figs. 4a and 4b, respectively. In Fig. 4a, $A$ is fixed at 0.2, and in Fig. 4b, $\Omega$ is fixed at 1.1. It can be seen from Fig. 4a that: first, near the natural frequency, i.e., $\Omega=1.1$, system (3) exhibits chaos, the smaller the damping is, the larger is the region of the parameter $\Omega$ having chaos; second, with the rotation frequency moving away from the natural frequency, the chaotic motion becomes a periodic oscillation; third, in the small damping coefficient range, there are periodic motion windows; the smaller the damping coefficient is, the smaller is the periodic window width; fourth, in a practical situation, period one is desirable if the oscillation is unavoidable, which means that the rotation speed should be set away from the natural frequency of the system. In addition, the small rotation speed corresponds to small oscillation amplitude.

From Fig. 4b, for fixed $\Omega=1.1$, we learn that: first, with the damping coefficient decreasing, the parameter range of the excitation amplitude, where chaos can be observed, becomes larger; second, there exists a periodic window between two chaotic parameter regions; third, with the damping coefficient decreasing, the chaotic parameter region becomes large. If the damping coefficient is large enough, chaos is eliminated.

Parameter space plot in Fig. 5 shows the different kinds of system responses when two parameters of the system are varied. It can be seen from Fig. 5 that: first, the system is in period one motion when the rotation frequency is less than the natural frequency; second, when the excitation amplitude is $A>0.2$, it is easier for the system to be in a higher period or chaos; third, there are periodic windows in the chaotic region.

From the parameter space plot in Fig. 5, we know that, if the exciting amplitude is less than $A<0.06$, the irregular swing can be avoided. This means that the irregular swing is disappeared if the eccentricity is small enough. And we also know that the irregular swing can be avoided by selecting the rotation speed less than $\Omega<1.07$, which give the helpful guide for the operator of the SCP to set the process parameter.

The largest Lyapunov exponent (LLE) of a dynamical system is a quantity that characterizes the average exponential separation between two phase trajectories that are initially close by. In the chaotic region, the LLE must be positive. In the following, we give the LLE variation versus the parameter variation in order to clearly see the relationship of the LLE and the dynamics of the system, the bifurcation diagram with the same parameter variation is also given.

In the first case, $A$ is fixed at 0.2 and the damping coefficient $c$ is fixed at 0.3, then the LLE variation and the corresponding bifurcation diagram versus $\Omega$ variation are given in Figs. 6a and 6b, respectively. In the second case, $\Omega$ is fixed at 1.1 and the damping coefficient $c$ is fixed at 0.1, then the LLE variation and the corresponding bifurcation diagram versus $A$ variation are given in Figs. 7a and 7b, respectively. From Figs. 6 and 7, we know that in the chaotic parameter region the LLE is positive.

\begin{figure*}
\centering
\subfigure{
  \includegraphics[width=2.5in]{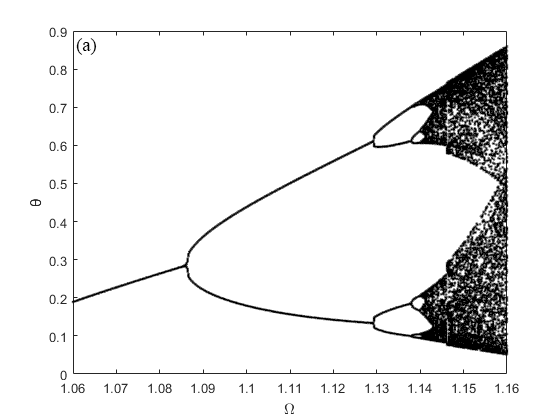}
}

\subfigure{
  \includegraphics[width=1.2in]{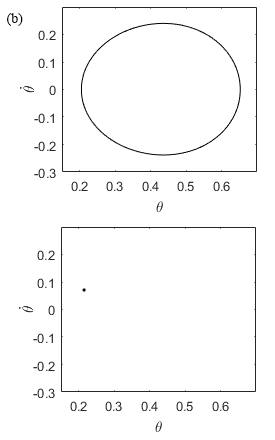}
}
\subfigure{
  \includegraphics[width=1.2in]{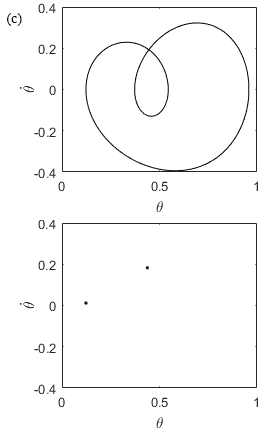}
}
\subfigure{
  \includegraphics[width=1.2in]{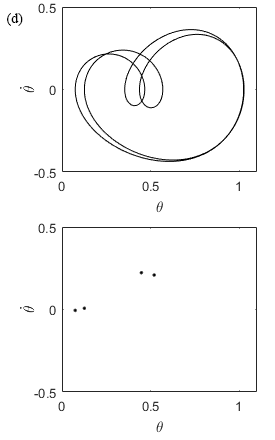}
}

\subfigure{
  \includegraphics[width=1.2in]{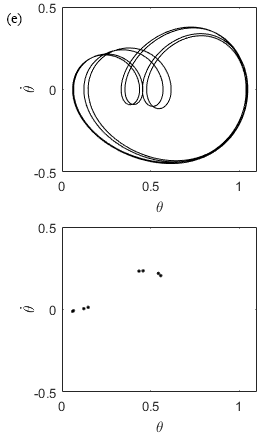}
}
\subfigure{
  \includegraphics[width=1.2in]{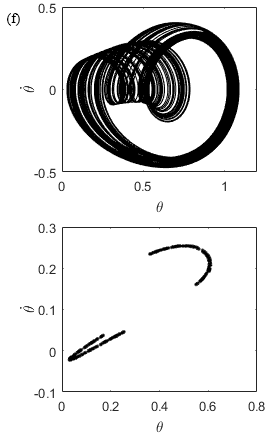}
}
\caption{(a) Local blow up bifurcation diagram shown in fig. 6(a) at $\Omega\in(1.06,1.16)$. (b) The phase trajectory and Poincar\'{e} section point of period-1 oscillation at $\Omega=1.068$ are shown in the upper panel and in the lower panel, respectively. (c) The phase trajectory and Poincar\'{e} section points of period-2 oscillation at $\Omega=1.115$ are shown in the upper panel and in the lower panel, respectively. (d) The phase trajectory and Poincar\'{e} section points of period-4 oscillation at $\Omega=1.133$ are shown in the upper panel and in the lower panel, respectively. (e) The phase trajectory and Poincar\'{e} section points of period-8 oscillation at $\Omega=1.139$ are shown in the upper panel and in the lower panel, respectively. (f) The phase trajectory and Poincar\'{e} section points of chaos at $\Omega=1.15$ are shown in the upper panel and in the lower panel, respectively.}
\end{figure*}

\begin{figure*}
\centering
\subfigure{
  \includegraphics[width=2.5in]{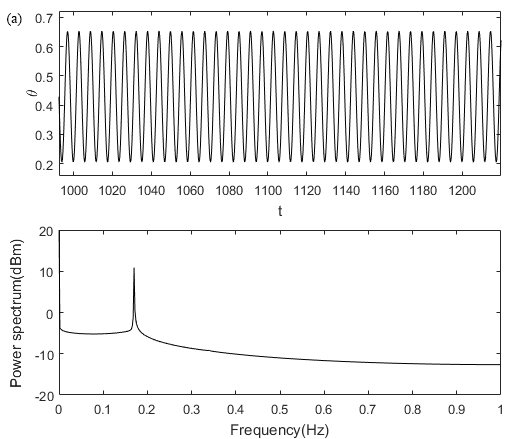}
}
\subfigure{
  \includegraphics[width=2.5in]{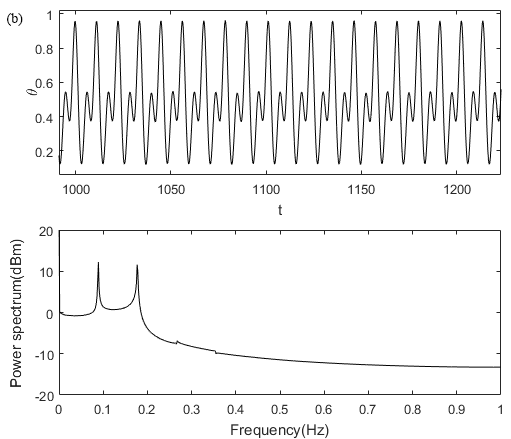}
}

\subfigure{
  \includegraphics[width=2.5in]{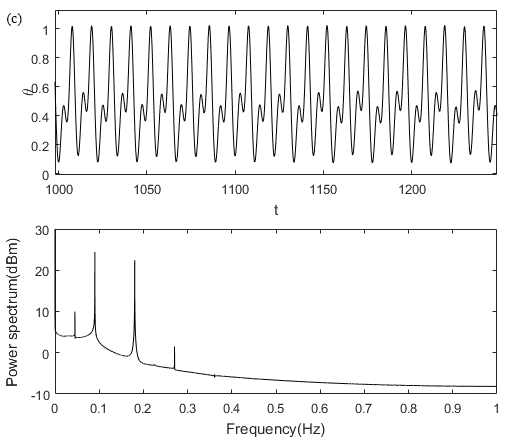}
}
\subfigure{
  \includegraphics[width=2.5in]{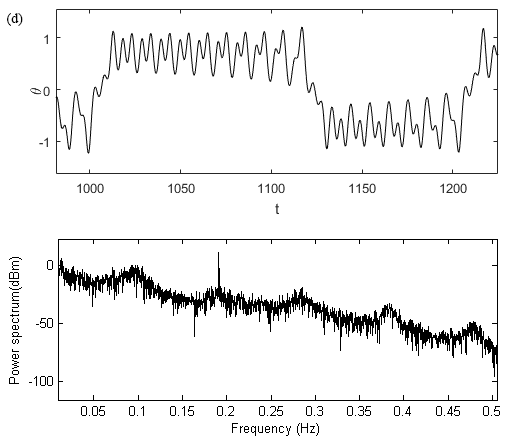}
}


\caption{(a) The time sequence and the corresponding power spectrum of period-1 oscillation at $\Omega=1.068$ are shown in the upper panel and in the lower panel, respectively. (b) The time sequence and the corresponding power spectrum of period-2 oscillation at $\Omega=1.115$ are shown in the upper panel and in the lower panel, respectively. (c) The time sequence and the corresponding power spectrum of period-4 oscillation at $\Omega=1.133$ are shown in the upper panel and in the lower panel, respectively. (d) The time sequence and the corresponding power spectrum of chaos at $\Omega=1.2$ are shown in the upper panel and in the lower panel, respectively.}
\end{figure*}

\begin{figure*}
\centering
  \includegraphics[width=3.25in]{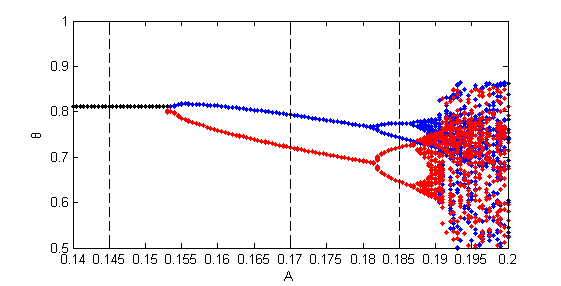}
\caption{An amplified window of Fig. 7(a) exhibiting symmetry-breaking bifurcation.}
\end{figure*}

\begin{figure*}
\centering
\subfigure{
  \includegraphics[width=1.2in]{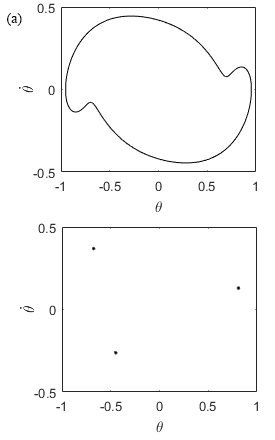}
}
\subfigure{
  \includegraphics[width=1.2in]{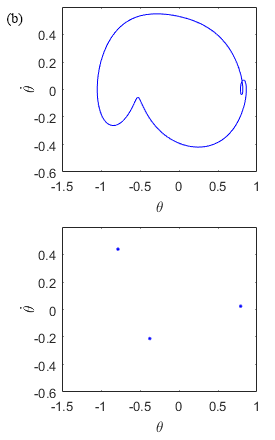}
}
\subfigure{
  \includegraphics[width=1.2in]{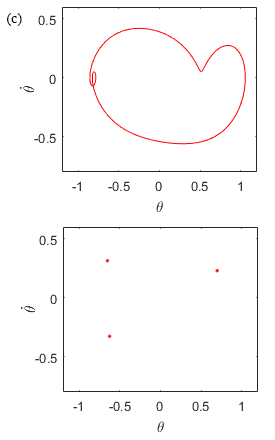}
}

\subfigure{
  \includegraphics[width=1.2in]{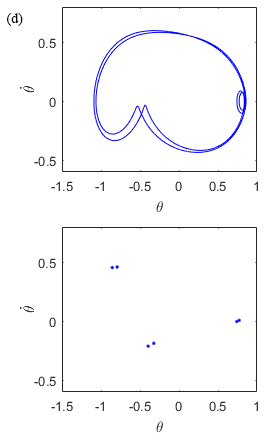}
}
\subfigure{
  \includegraphics[width=1.2in]{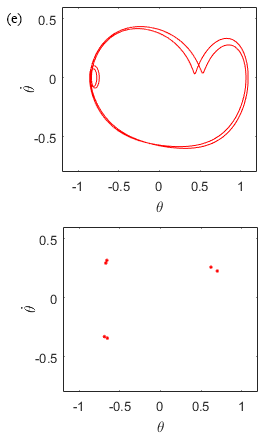}
}
\caption{The phase portraits and Poincar\'{e} sections for three different A values in the symmetry-breaking bifurcation in Fig. 10. Subplot (a) is the phase trajectory and Poincar\'{e} section for A=0.145. Subplots (b) and (c) are the phase trajectories and Poincar\'{e} sections for A=0.17 at different initial values; Subplots (d) and (e) are the phase trajectories and Poincar\'{e} sections for A=0.185 at different initial conditions.}
\end{figure*}

\subsection{Routes to Chaos}
\paragraph{A. Period doubling bifurcation} \hspace{0pt}\\
Period doubling bifurcation is one of the most common routes from periodic motion to chaos. From Figs. 4 to 7, we can observe many examples of this route. To clearly see this point, we show the blow up bifurcation diagram within the range $\Omega\in(1.06,1.16)$ in Fig. 6a, shown in Fig. 8a. From Fig. 8a, we know that when $\Omega\in(1.06,1.087)$, period one is observed; when $\Omega\in(1.087,1.129)$, period two is observed; when $\Omega\in(1.129,1.137)$, period four is observed; afterwards, period eight and then chaos are observed. To see the different dynamical behaviors, we give the phase trajectories and the corresponding Poincar\'{e} sections for $\Omega=1.068$, $\Omega=1.115$, $\Omega=1.133$, $\Omega=1.139$, and $\Omega=1.15$ in the upper panel and in the lower panel of Figs. $8b\sim 8f$, respectively.

To see the time sequence of different periods and chaos, we give the time sequence for $\Omega=1.068$, $\Omega=1.115$, $\Omega=1.133$, and $\Omega=1.2$ and their corresponding power spectrum in the upper panel and lower panel of Figs. $9a\sim 9d$, respectively. From Fig. 9, we surmise that the periodic oscillations are consistent with the corresponding power spectrum. In addition, chaos has a wide spectrum.

\paragraph{B. Symmetry-breaking bifurcation} \hspace{0pt}\\
In Fig. 8a, a periodic window occurs when $A\in(0.125,0.193)$. We notice that this periodic window are of period-3. In this interval, symmetry-breaking bifurcation takes place at $A=0.1535$. To exhibit how the system enters chaos through symmetry-breaking route, a detailed bifurcation diagram is given in Fig. 10. In fact, Fig. 10 shows a blow up of a local region in the bifurcation diagram of Fig. 7a. The black points in Fig. 10 represent the symmetric period-3 oscillations, while the blue and red points correspond to the two asymmetric period-3 solutions. For $A>0.182$, the asymmetric solutions simultaneously undergo period doubling. Following the bifurcation cascades, finally chaos occurs. The system state is asymptotic to the blue asymmetric or the red asymmetric depending on the initial conditions [21].

\begin{figure*}[!ht]
\subfigure{
  \centering
  \includegraphics[width=3.25in]{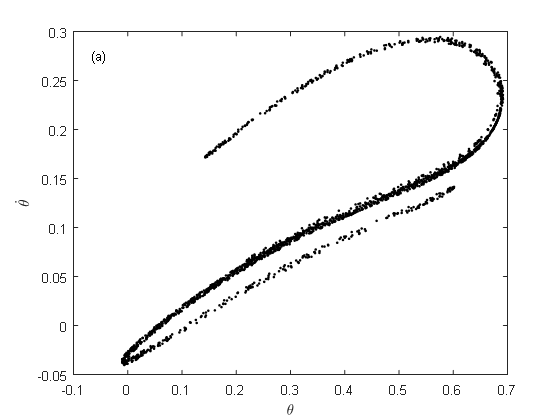}
}
\subfigure{
  \centering
  \includegraphics[width=3.25in]{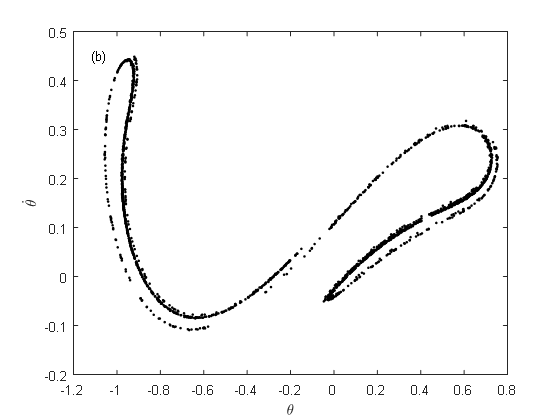}
}
\caption{(a) The Poincar\'{e} section of a small chaotic attractor at $\Omega=1.17$ in the bifurcation diagram of Fig. 6(a). (b) The Poincar\'{e} section of a large chaotic attractor at $\Omega=1.18$. in the bifurcation diagram of Fig. 6(a).}
\end{figure*}
\begin{figure*}
\subfigure{
  \centering
  \includegraphics[width=3.25in]{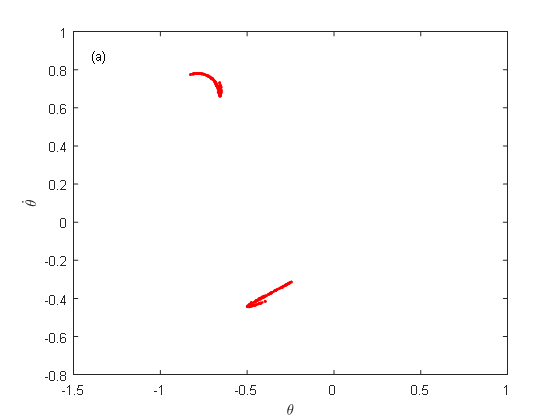}
}
\subfigure{
  \centering
  \includegraphics[width=3.25in]{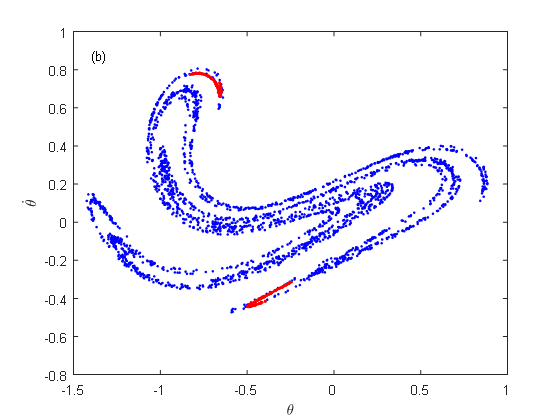}
}
\caption{(a) The Poincar\'{e} section of two isolated weak chaotic attractors located at $A=0.3$ in Fig. 7(a). (b) The Poincare section of a single strong chaotic attractor located at $A=0.301$ in Fig. 7(a).}
\end{figure*}

\paragraph{C. Interior crisis} \hspace{0pt}\\
In this subsection, we show that there is another route to chaos, namely, interior crisis. In Fig. 6a, it can be observed that, at $\Omega\approx1.18$, a small chaotic region suddenly enlarges into a larger one, which is a typical interior crisis phenomenon [22]. Figure 12 shows this change in the Poincar\'{e} section. This interior crisis, which is a type of global bifurcation, is another route to chaos when the parameter of the FSLR system varies.

The interior crisis can also be observed in Fig. 7a, where we can see that, for $A=0.3$, there exist two isolated small chaotic region whose Pioncar\'{e} section is shown in Fig. 13a, while, for $A=0.301$, there is one single large chaos region whose Pioncar\'{e} section is shown in Fig. 13b. As parameter $A$ passes the critical value, the size of the attractor is suddenly enlarged. The new blue Poincar\'{e} section points include the old red ones and new incremental blue points. This is a typical interior crisis phenomenon.

\section{Conclusions}
In this paper, the dynamical model of the FSRL system of SCP is established based on the working principle of the FSRL system. The Melnikov method, the bifurcation diagram, the Lyapunov exponents, phase trajectories, Poincar\'{e} sections and power spectra have been used to investigate the dynamical behaviors of the system. We learn from the analysis of this paper that: first, the rotation speed, i.e., the excitation frequency, the amplitude of excitation depending on the degree of eccentricity, and the damping coefficient affect the dynamical behaviors of the system; second, depending on different parameters, the system demonstrates a tremendous variety of different dynamical behaviors, including period-1, period-2, ..., and chaos; third, when the excitation frequency is close to the natural frequency of the system, complex behaviors, including high period and chaos occurs, which is consistent with the practical observations from industrial plants.We have shown three routes to chaos of the FSRL system, namely, the period doubling bifurcation, symmetry-breaking bifurcation, and the interior crisis route.

The complex dynamic characteristics of the system investigated in this paper explain the irregular swing phenomenon observed in the practical plants, and it provides a theoretical basis for eliminating the unexpected swing phenomenon of the FSRL system in the SCP using Cz method. Designing the eccentricity to be zero for the mechanical engineer is too challenge task to accomplish, therefore, designing an active controller to control the swing is a more feasible and adaptive method to deal with the swing problem, which will be given in the future paper.

\end{document}